\documentclass[12pt,preprint]{emulateapj}

\begin{document}

\shorttitle{X-rays from PSR J0030+0451}
\shortauthors{Bogdanov \& Grindlay}

\title{Deep XMM-Newton Spectroscopic and Timing Observations \\ of the
  Isolated Radio Millisecond Pulsar PSR J0030+0451}

\author{Slavko Bogdanov\altaffilmark{1,2,3} and Jonathan E. Grindlay\altaffilmark{1}} 

\altaffiltext{1}{Harvard-Smithsonian Center for Astrophysics, 60 Garden Street,
Cambridge, MA 02138; sbogdanov@cfa.harvard.edu, josh@cfa.harvard.edu}

\altaffiltext{2}{Department of Physics, McGill University, 3600 University Street, Montreal, QC H3A 2T8, Canada}

\altaffiltext{3}{Canadian Institute for Advanced Research Junior Fellow}

\begin{abstract}
  We present deep \textit{XMM-Newton} EPIC spectroscopic and timing
  X-ray observations of the nearby solitary radio millisecond pulsar,
  PSR J0030+0451.  Its emission spectrum in the 0.1--10 keV range is
  found to be remarkably similar to that of the nearest and best
  studied millisecond pulsar, PSR J0437--4715, being well described by
  a predominantly thermal two-temperature model plus a faint hard tail
  evident above $\sim$2 keV.  The pulsed emission in the 0.3--2 keV
  band is characterized by two broad pulses with pulsed fraction
  $\sim$60-70\%, consistent with a mostly thermal origin of the
  X-rays only if the surface polar cap radiation is from a
  light-element atmosphere. Modeling of the thermal pulses permits us
  to place constraints on the neutron star radius of $R>10.7$ (95\%
  confidence) and $R > 10.4$ km (at 99.9\% confidence) for $M=1.4$
  M$_{\odot}$.
\end{abstract}


\section{INTRODUCTION}

PSR J0030+0451 is one of the nearest known rotation-powered
``recycled'' millisecond pulsars (MSPs) in the field of the Galaxy
($D=300\pm90$ pc; Lommen et al. 2006), with a spin period $P=4.87$ ms
and intrinsic spindown rate $\dot{P}\equiv {\rm d}P/{\rm d}t =
1.0\times10^{-20}$ s s$^{-1}$, implying a surface dipole magnetic
field strength $B\approx2.7\times10^8$ Gauss, a characteristic age
$\tau\approx7.8$ Gyr, and spin-down luminosity
$\dot{E}\approx3\times10^{33}$ ergs s$^{-1}$.  This solitary MSP was
discovered at radio frequencies in the Arecibo drift scan survey
\citep{Lom00}. Recently, a firm detection of PSR J0030+0451 at
$\gamma$-ray energies by the \textit{Fermi} Large Area Telescope was
reported by \citet{Abdo09}. In X-rays, PSR J0030+0451 has been detected
with \textit{ROSAT} PSPC \citep{Beck00} and \textit{XMM-Newton}
\citep{Beck02}. Despite the limited photon statistics, these
observations clearly showed two distinct broad pulses. In addition,
the \textit{XMM-Newton} observation revealed a relatively soft 0.3--2
keV spectrum, qualitatively similar to that of PSR J0437--4715
\citep{Zavlin06}.

X-ray observations have detected a number of rotation-powered MSPs
that exhibit soft, presumably thermal, emission
\citep{Zavlin06,Bog06a,Zavlin07}. This radiation likely originates
from the pulsar magnetic polar caps that are heated by energetic
particles from the pulsar magnetosphere \citep[see,
e.g.,][]{Hard02}. As this heat is confined to a small portion of the
NS, study of the X-ray properties of MSPs could offer insight into key
NS properties that are inaccessible by other observational means
(e.g., radio pulse timing) such as the radiative properties of the NS
surface, magnetic field geometry, and NS compactness ($R/R_S$, where
$R_S=2GM/c^2$ and $R$ and $M$ are the stellar mass and radius). As
shown by \citet{Pavlov97}, \citet{Zavlin98}, and \citet{Bog07}, a
model of polar cap thermal emission from an optically-thick hydrogen
(H) atmosphere provides a good description of the X-ray pulse profiles
of PSR J0437--4715, the nearest known MSP. On the other hand, a
blackbody model is inconsistent with the pulsed X-ray
emission. Furthermore, there is compelling evidence for a magnetic
dipole axis offset from the NS center \citep{Bog07}. Finally, the
compactness of PSR J0437--4715 is constrained to be $R/R_S>1.6$
(99.9\% confidence), which for the current best mass measurement
\citep[$1.76$ M$_{\odot}$, see][]{Verb08} implies $R>8.3$ km. \citet{Bog08}
have shown, using short archival \textit{XMM-Newton} observations of
PSRs J0030+0451 and J2124-3358, that even with relatively crude photon
statistics, it is possible to set limits on $M/R$. In particular,
assuming 1.4 M$_{\odot}$, the stellar radius is constrained to be $R >
9.4$ km and $R > 7.8$ km (68\% confidence) for each pulsar,
respectively. Thus, realistic modeling of X-ray data from MSPs appears
to be a promising approach towards revealing the structure of NSs and
warrants further X-ray observations of nearby MSPs. As one of the
nearest known recycled pulsars, PSR J0030+0451 is well suited for such
an investigation.

In this paper, we present an analysis of deep \textit{XMM-Newton}
spectroscopic and timing observations of PSR J0030+0451. These
observations provide new insight into the X-ray properties of MSPs and
the structure of neutron stars.  The work is presented as follows. In
\S 2 we summarize the observations and data reduction procedure. In \S
3 we briefly examine the X-ray images. In \S4 we constrain the
spectral properties of the pulsar, while in \S 5 we conduct a timing
analysis. In \S 6 we attempt to model the pulsed emission from the
MSP. We offer conclusions in \S 7.

\section{OBSERVATION AND DATA REDUCTION}

PSR J0030+0451 was targeted by \textit{XMM-Newton} between 12 and 14
December 2007 (observation ID 050229). The observation was carried
out in a single uninterrupted 130-kilosecond exposure, corresponding
to the entire usable observing time of \textit{XMM-Newton} revolution
1467.  The European Photon Imaging Camera (EPIC) MOS1 and MOS2
instruments were configured for full imaging mode. The EPIC pn was
configured for fast timing mode, in which only CCD 4 is active,
allowing 30 $\mu$s time resolution at the expense of one imaging
dimension. For all three instruments, the thin optical filter was
used.  Due to the faint nature of J0030+0451, the dispersed Reflection
Grating Spectrometer (RGS) data provide no useful spectral or timing
information. Therefore, the grating data were not used in our
analysis.

%
%   FIGURE 1
%
\begin{figure}[!t]
\begin{center}
\includegraphics[width=0.47\textwidth]{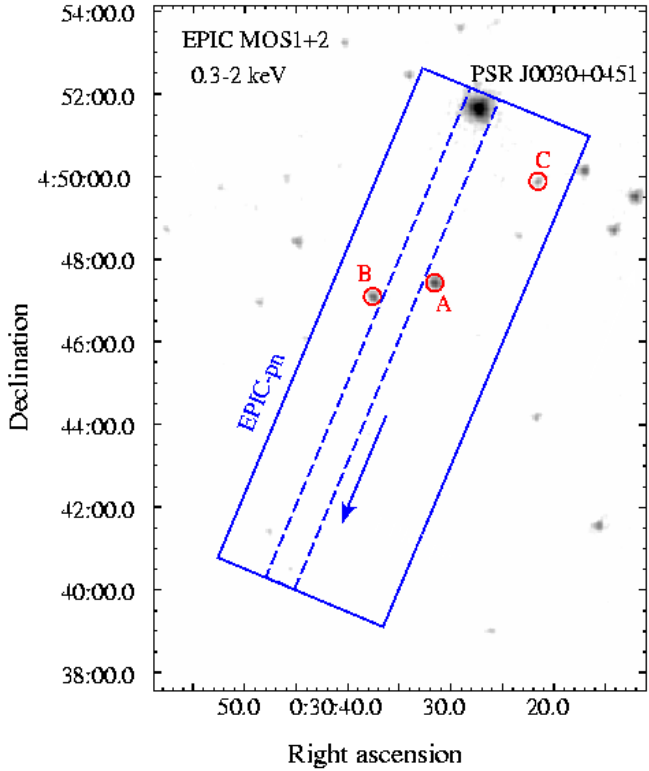}
\end{center}
\caption{\textit{XMM-Newton} MOS1 and MOS2 mosaic image of the field
  near PSR J0030+0451 in the 0.3--2 keV band. The rectangle shows the
  relative position and orientation of the active EPIC pn chip used in
  this observation, with the timing redout direction indicated by the
  arrow. The dashed lines correspond to the EPIC pn detector columns
  used to extract the source counts from the pulsar while the circles
  mark the positions of sources that fall within the EPIC pn field of view.}
\end{figure}

The data reduction, imaging, and timing analyses were performed using
the SAS\footnote{The \textit{XMM-Newton} SAS is developed and
  maintained by the Science Operations Centre at the European Space
  Astronomy Centre and the Survey Science Centre at the University of
  Leicester.} 8.0.1 and FTOOLS\footnote{Available at
  http://heasarc.gsfc.nasa.gov/ftools/} 6.6.1 software packages, while
the spectral analysis was conducted in XSPEC\footnote{Available at
  http://heasarc.nasa.gov/docs/xanadu/xspec/} 12.5.0.  The raw MOS and
pn datasets were first reprocessed using the SAS {\tt emchain} and
{\tt epchain} pipelines, respectively, and subsequently screened for
instances of high background proton flares.  A total of $\sim$36 ks of
high-background data were discarded, resulting in 92.5, 92.1, and 92.1
ks of usable exposure time for MOS1, MOS2, and pn,
respectively. Finally, the appropriate pattern, pulse invariant, and
flag filters were applied to produce data suitable for spectral and
timing analyses.

For the spectroscopic analysis, source counts in the MOS 1 and 2 data
were extracted from circles of radius 60'' centered on the radio MSP
position, which enclose $\sim$90\% of the total energy at 1.5 keV.
The background was taken from three source-free regions surrounding
the pulsar. For the spectroscopic and timing analyses of the EPIC pn
dataset, the source counts were obtained from RAWX detector columns
31-41 (inclusive), equivalent to a radius of 22.5'' in the RAWX
direction, which encloses $\sim$74\% of the total energy for 0.3--2
keV. This relatively narrow region was chosen in order to avoid
contamination from sources A and B as well as to minimize the large
background level, which dominates beyond $\sim$20'' from the source
position, arising due to the 1-dimensional imaging mode used for the
pn detector. To ensure a reliable estimate, the background was taken
from three source-free regions.

%
%   FIGURE 2
%
\begin{figure}[!t]
\begin{center}
\includegraphics[width=0.47\textwidth]{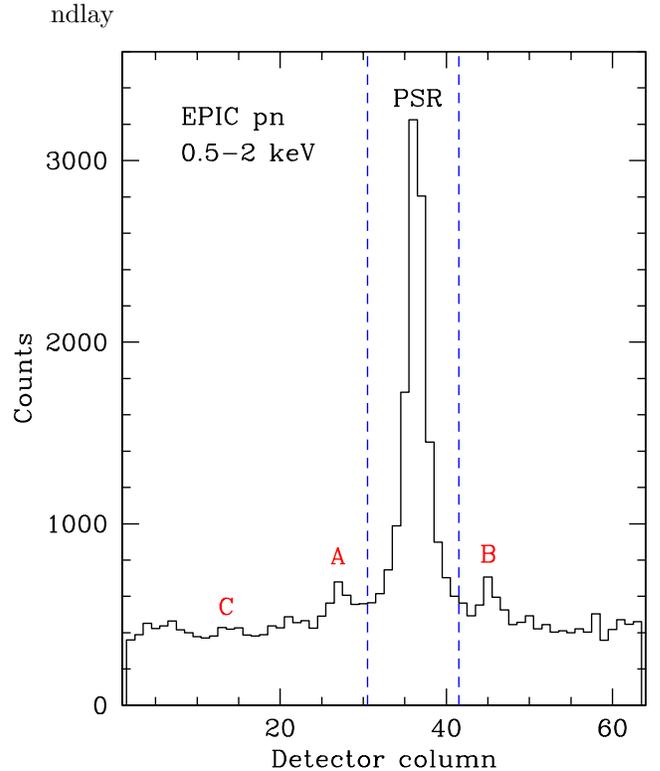}
\end{center}
\caption{The distribution of counts from \textit{XMM-Newton} EPIC pn
  along the imaging dimension of the fast timing mode in the 0.5--2
  keV band. The pair of dashed lines mark the detector columns
  that the pulsar counts were extracted from. The labels show the
  positions of the sources from Figure 1.}
\end{figure}

\section{IMAGING ANALYSIS}

Figure 1 shows the 0.3--2 keV coadded image from EPIC MOS1 and MOS2 of
the field in the vicinity of PSR J0030+0451, as well as the relative
orientation and readout direction of the EPIC pn detector. It is clear
that the pulsar is the brightest source in the image. Due to the
favorable telescope roll angle during the observation, the EPIC pn
detector columns containing the bulk of counts from the pulsar are
effectively free of contamination from other sources (Fig.~2). Sources
A and B combined contribute with only a few counts ($\lesssim$0.1\% of
the total) in the source extraction colums of the pulsar.

In the MOS1 and 2 images, the source count distribution around the
radio position of the pulsar is fully consistent with that of a point
source. Thus, we find no indication of any diffuse extended emission
that could arise due to a bow shock, as seen in PSR J2124--3358
\citep{Hui06}. This is not surprising given that for the combination
of spin-down luminosity ($\dot{E}\approx 3\times 10^{33}$ ergs
s$^{-1}$) and remarkably low transverse space velocity \citep[$8-17$
km s$^{-1}$, see][]{Lom06} for this pulsar, no detectable diffuse X-ray
emission is expected \citep[see, e.g., Fig.~1 in][]{Chat02}.

\begin{deluxetable*}{lccccccccl}[]
\tabletypesize{\small} 
\tablecolumns{10} 
\tablewidth{0pc}
\tablecaption{Best fit spectral models and unabsorbed fluxes for PSR J0030+0451.}

\tablehead{ \colhead{} & \colhead{$N_{\rm H}$} &
\colhead{$T_{\rm eff,1}$} & \colhead{$R_{\rm eff,1}$\tablenotemark{b}} & \colhead{$T_{\rm eff,2}$} & \colhead{$R_{\rm eff,2}$\tablenotemark{b}} &  \colhead{ } &  & \colhead{$F_{\rm X}$\tablenotemark{c}} & \colhead{$\chi^2_{\nu}$/dof} \\
\colhead{Model\tablenotemark{a}} & \colhead{($10^{20}$ cm$^{-2}$)} & \colhead{($10^6$ K)} & \colhead{(km)} & \colhead{($10^6$ K)} & \colhead{(km)} &   & \colhead{} & \colhead{(0.1--10 keV)}  &  }

\startdata
BB($\times$2)	&	$0.90^{+0.3}_{-0.3}$	&	$3.05^{+0.15}_{-0.16}$	&	$0.05^{+0.03}_{-0.03}$	&	$1.45^{+0.07}_{-0.08}$	&	$0.28^{+0.13}_{-0.10}$	&	-	&	-	&	$3.1\pm0.1$	&	$1.21/299$	\\

Hatm($\times$2)	&	$2.0^{+0.4}_{-0.3}$	&	$1.71^{+0.16}_{-0.15}$	&	$0.25^{+0.04}_{-0.03}$	&	$0.68^{+0.07}_{-0.07}$	&	$2.2^{+0.7}_{-0.6}$	&	-	&	-	&	$3.7\pm0.1$	&	$1.16/299$	\\

\hline
\colhead{} &  &
\colhead{} & \colhead{} & \colhead{} & \colhead{} &  \colhead{$\Gamma$} &   &  &  \\
\hline
BB+PL	&	$4.0^{+0.4}_{-0.4}$	&	$2.11^{+0.05}_{-0.04}$	&	$0.11^{+0.04}_{-0.04}$	&	-	&	-	&	$3.09^{+0.08}_{-0.08}$	& - &	$8.6\pm0.1(7.1\pm0.1)$	&	$1.12/299$	\\

Hatm+PL	&	$3.4^{+0.4}_{-0.4}$	&	$1.06^{+0.03}_{-0.03}$	&	$0.86^{+0.37}_{-0.30}$ &	-	& - &  $3.06^{+0.09}_{-0.09}$	& - &	$6.9\pm0.1(4.8\pm0.1)$	&	$1.11/299$	\\

BB$(\times$2)+PL	&	$3.8^{+0.5}_{-0.7}$	&	$2.34^{+0.86}_{-0.36}$	&	$0.07^{+0.05}_{-0.04}$	&	$1.75^{+0.33}_{-0.24}$	&	$0.11^{+0.18}_{-0.13}$	&	$3.08^{+0.09}_{-0.09}$ &   -	&	$8.2\pm0.2(6.6\pm0.2)$	&	$1.10/297$	\\

	&	$1.5^{+0.4}_{-0.3}$	&	$2.51^{+0.11}_{-0.11}$	&	$0.08^{+0.02}_{-0.02}$	&	$1.18^{+0.08}_{-0.08}$	&	$0.34^{+0.20}_{-0.17}$	&  $1.4$ & - &	$3.7\pm0.2(0.5\pm0.1)$	&	$1.13/298$	\\

Hatm($\times$2)+PL	&	$1.9^{+0.7}_{-0.6}$	&	$1.42^{+0.12}_{-0.11}$	&	$0.37^{+0.02}_{-0.01}$	&	$0.69^{+0.08}_{-0.08}$	&	$1.9^{+0.19}_{-0.16}$	&	$2.0^{+0.2}_{-0.2}$ &  -	&	$3.9\pm0.4(0.6\pm0.1)$	&	$1.12/297$	\\

	&	$1.8^{+0.2}_{-0.5}$	&	$1.51^{+0.08}_{-0.07}$	&	$0.3^{+0.2}_{-0.2}$	&	$0.70^{+0.08}_{-0.07}$	&	$1.97^{+0.41}_{-0.38}$ &	$1.4$	& - &	$3.7\pm0.3(0.4\pm0.1)$	&	$1.11/298$	\\

\hline
\colhead{} &  & \colhead{} & \colhead{} & \colhead{} & \colhead{} &  \colhead{$kT_{e}$} & \colhead{$\tau$}  & \colhead{} & \colhead{} \\
\colhead{} &  & \colhead{} & \colhead{} & \colhead{} & \colhead{} &  \colhead{(keV)} &   & \colhead{} & \colhead{} \\
\hline
Compbb($\times$2)	&	$1.8^{+0.5}_{-0.4}$	&	$2.28^{+0.10}_{-0.10}$	&	$0.11^{+0.06}_{-0.05}$	&	$1.03^{+0.12}_{-0.09}$	&	$0.38^{+0.27}_{-0.19}$	&	$100$	&	$0.12\pm0.01$	&	$3.8\pm0.3$	&	$1.10/298$	\\

\hline
\colhead{} &  &
\colhead{} & \colhead{} & \colhead{} & \colhead{} &  \colhead{$T_{\rm eff,3}$} & $R_{\rm eff,3}$  &    &  \\
\colhead{} &  &
\colhead{} & \colhead{} & \colhead{} & \colhead{} &  \colhead{($10^6$ K)} & (km)             &   &  \\
\hline

BB($\times$3)	&	$1.9^{+0.6}_{-0.5}$	&	$6.98^{+1.40}_{-1.21}$	&	$0.004^{+0.005}_{-0.003}$	&	$2.27^{+0.13}_{-0.14}$	&	$0.12^{+0.07}_{-0.06}$	&	$1.01^{+0.12}_{-0.14}$	&	$0.46^{+0.48}_{-0.29}$	&		$3.7\pm0.3$	& $1.10/297$
		\\
Hatm($\times$3)	&	$2.2^{+0.6}_{-0.6}$	&	$3.36^{+0.84}_{-0.78}$	&	$0.03^{+0.06}_{-0.03}$	&	$1.05^{+0.07}_{-0.09}$	&	$0.95^{+0.8}_{-0.8}$	&	$0.40^{+0.06}_{-0.07}$	&	$4.3^{+2.8}_{-2.1}$	&	$4.0\pm0.4$	&	$1.11/297$	\\

\enddata

\tablenotetext{a}{PL is a powerlaw, BB a blackbody, and Hatm a H
  atmosphere model. All uncertainties quoted are 1$\sigma$.}

\tablenotetext{b}{$R_{\rm eff}$ calculated assuming a distance of 300
  pc. For the H atmosphere model, the numbers quoted represent
  deprojected and redshift-corrected effective radii of one polar cap
  of a neutron star with $M=1.4$ M$_{\odot}$, $R=12$ km,
  $\alpha=70^{\circ}$, and $\zeta=80^{\circ}$ (see text for details).}

\tablenotetext{c}{Unabsorbed X-ray flux (0.1--10 keV) in units of
  $10^{-13}$ ergs cm$^{-2}$ s$^{-1}$. The value in parentheses
  represents the flux contribution of the powerlaw component.}

\end{deluxetable*}

\section{SPECTROSCOPIC ANALYSIS}

To facilitate the spectral fitting of the MOS1 and MOS2 data, the
extracted counts in the 0.1--10 keV band were grouped so as to ensure
at least 30 counts per bin. For the pn data the counts in the 0.3--3
keV were grouped with at least 150 counts per bin. Due to the
overwhealming background level, no useful spectral information is
present above 3 keV in the pn data. In addition, photons in the
0.42--0.5 keV range were not used in order to eliminate an
instrumental noise artifact specific to the fast timing mode of the pn
detector.

As found in previous studies of PSR J0030+0451
\citep{Beck02,Bog08,Zavlin07}, the continuum X-ray emission from this
pulsar cannot be adequately described by a single component emission
model. Consequently, in the spectral fits we apply the following
plausible multi-component models: i) two thermal, ii) a single thermal
plus non-thermal, iii) two Comptonized thermal iv) two thermal plus
one non-thermal, and v) three thermal components. For the thermal
components we consider both a blackbody model and the unmagnetized
($B=0$) H atmosphere model first presented in \citet{McC04} and
included as a private model in {\tt XSPEC}. The assumption of $B=0$ is
appropriate since for typical MSP magnetic fields ($\sim$$10^{8-9}$ G)
the magnetic field does not affect the radiative opacities and
equation of state of the atmosphere \citep[see][for
details]{Rom87,Zavlin96,McC04}.

It has been found that blackbody fits to MSP spectra result in
effective emission areas that are an order of magnitude smaller than
the expected polar cap area, $R_{pc}=(2\pi R/cP)^{1/2}R$, while for a
H atmosphere they are found to be comparable
\citep{Beck02,Zavlin02,Bog06a,Zavlin06}.  The discrepancy could, in
principle, arise due to non-uniform heating of the polar caps. More
plausably, it suggests that the surface of the polar caps is covered
by a light-element atmosphere. As the spectrum of a H atmosphere model
is harder than that of a blackbody for the same effective temperature,
a blackbody model fitted to a H atmosphere continuum would yield a
higher inferred temperature and, as a result, a smaller effective
area.  Nonetheless, for the sake of completeness and easier comparison
with previous X-ray studies of MSPs, we have considered blackbody
models in our spectral fits as well.

%
%  FIGURE 3
%
\begin{figure}[t]

\includegraphics[width=0.47\textwidth]{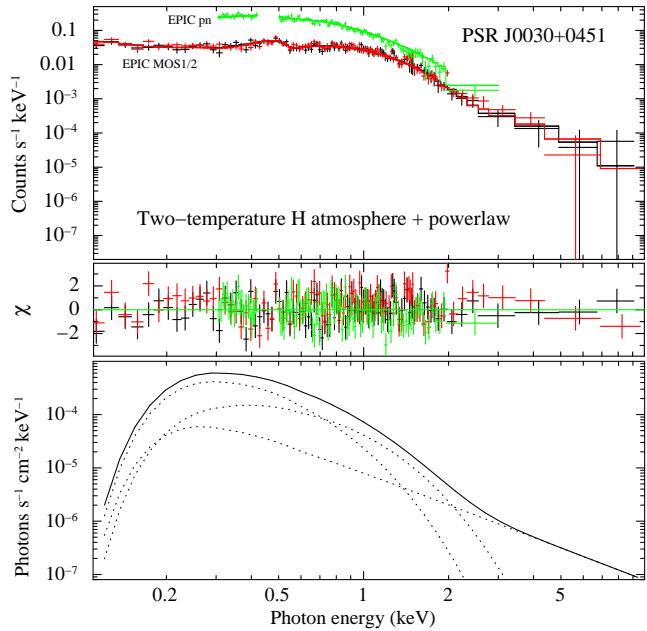}
\caption{The \textit{XMM-Newton} EPIC X-ray spectra of PSR J0030+0451
  fitted with a two-temperature H-atmosphere plus powerlaw model. The
  middle panel shows the fit residuals while the bottom panel shows
  the best fit model. See Table 1 for best fit parameters.}
%\end{center}
\end{figure}

\citet{Zavlin02}, \citet{Zavlin06}, and \citet{Bog07} have noted that
due to the energy-dependent limb darkening effect for the atmosphere
model, even for phase-integrated spectra it is necessary to take into
account the rotation of the star as well as the relative orientation
of the spin axis, hot spots, and the line of sight to the observer in
order to ensure a reliable measurement of the temperature and emission
radius \citep[see, in particular, Fig.~2 of][]{Bog07}. For this
reason, we employ the spectral model from \citet{Bog07}. This model
allows us to specify the angles between the spin and magnetic axes
($\alpha$) and the spin axis and the line of sight ($\zeta$) as well
as the mass and radius of the pulsar in the spectral fits. Thus, the
normalization of the thermal component(s) yields the true (deprojected
and redshift-corrected) effective area of the emission region(s). In
addition, the quoted best-fit temperatures represent
redshift-corrected values, i.e. as they would be measured at the NS
surface.  For the calculation of all effective radii, we consider the
parallax distance $D=300$ pc.  In the case of the H atmosphere model,
we assume a neutron star with $M=1.4$ M$_{\odot}$ and $R=12$ km. Based
on the X-ray pulse profile morphology (discussed in \S5) we also fix
$\alpha=70^{\circ}$ and $\zeta=80^{\circ}$ since this combination of
angles can reproduce the X-ray pulse profile shape for $M=1.4$
M$_{\odot}$ and $R=12$ km.  All uncertainties quoted below are given
at a 1$\sigma$ level for one interesting parameter. The various
spectral models and their best fit parameters are summarized in Table
1.

\subsection{Two-Temperature Thermal Spectrum}

First, we consider the two-temperature thermal model that was found to
provide a good fit to the archival \textit{XMM-Newton} data of PSR
J0030+0451 \citep{Beck02,Bog08}. This model describes the new spectrum
well up to $\sim$3 keV but cannot fully account for the excess flux at
higher energies. Thus, it is necessary to invoke alternative emission
models.

\subsection{Thermal Plus Non-thermal Spectrum}

Next, we apply a composite thermal plus non-thermal (powerlaw) model.
As evident from Table 1, for either a blackbody or H atmosphere
thermal component the model yields statistically acceptable
fits. However, the derived values of $N_{\rm H}$ are significantly
larger than those inferred from independent measurements. In
particular, the total H column density through the Galaxy along the
line of sight to the pulsar is only $N_{\rm H}\approx3\times10^{20}$
cm$^{-2}$ \citep{Dicke90}.  On these grounds, we deem the validity of
the single temparature thermal plus non-thermal model doubtful.

\subsection{Comptonized Thermal Spectrum}

\citet{Bog06b} have proposed that the hard tail seen in the X-ray
spectrum of PSR J0437--4715 may be produced by inverse Compton
scattering (ICS) of the soft thermal X-rays by relativistic $e^{\pm}$
of low optical depth ($\tau \ll 1$). Such particles are likely present
in the pulsar magnetosphere above the polar caps.  In the spectral
fits of PSR J0030+0451, we employ the {\tt compbb} Comptonized
blackbody model \citep{Nish86}, which includes two additional
parameters: a scattering particle temperature $kT_e$ and optical depth
$\tau$. The model assumes a plane-parallel, semi-infinite scattering
medium with the source of thermal photons at the bottom. The formalism
employed is valid between $kT\lesssim kT_e\le 150$ keV. For the two
thermal components we link the $\tau$ parameter as both would likely
be scattered by the same population of particles.  For a choice of
$kT_e=100$ keV, the best fit parameters are $N_{\rm
  H}=1.8^{+0.4}_{-0.4}\times10^{20}$ cm$^{-2}$, $T_{\rm
  eff,1}=(2.28^{+0.10}_{-0.10})\times 10^6$ K, $R_{\rm
  eff,1}=0.11^{+0.06}_{-0.05}$ km, $T_{\rm
  eff,2}=(1.03^{+0.12}_{-0.09})\times 10^6$ K,
$R_2=0.38^{+0.27}_{-0.19}$ km, $\tau=0.12\pm0.01$ with
$\chi^2_{\nu}=1.10$ for 298 degrees of freedom.  The implied
unabsorbed flux is $3.8\times 10^{-13}$ ergs cm$^{-2}$ s$^{-1}$
(0.1--10 keV). As is generally the case for ICS, $kT_e$ and $\tau$ are
strongly correlated so acceptible fits were obtained for a wide range
of these parameters. Since we expect $kT \ll kT_e$, the choice of
$kT_e$ as well as the actual energy distribution of the scattering
particles do not significantly affect the X-ray spectral fits. Thus,
the ICS model appears to provide a good description of the spectrum of
PSR J0030+0451 as well.

\subsection{Two-Temperature Thermal Plus Powerlaw Spectrum}

\citet{Zavlin02} and \citet{Zavlin06} have found the continuum X-ray
emission from PSR J0437--4715 to be well described by a
two-temperature thermal and a single powerlaw components. As expected
based on the similar spin properties of the two pulsars, the same
model provides a good description of the X-ray spectrum of PSR
J0030+0451. Formally, an F-test indicates a $99.9997$\% probability
that the addition of the powerlaw component to the two-temperature
thermal model is required.  As with the single temperature composite
model, the derived best fit H column density for the two blackbody
plus powerlaw model with $\Gamma$ unconstrained are in excess of the
upper limit $N_{\rm H}=3\times10^{20}$ cm$^{-2}$ and yield a rather
steep spectral photon index ($\Gamma\sim3$). On the other hand, the
two-temperature H atmosphere plus powerlaw model (Fig.~3) results in a
more plausible value of $N_{\rm H}$ and a value of $\Gamma$ very
similar to that found for PSR J0437--4715 ($\Gamma\approx2$).  We note
that due to the poor photon statistics above $\sim$4 keV,
statistically acceptable fits are obtained for a wide range of photon
indices (1.0--3.5 at 3$\sigma$ confidence) although for
$\Gamma\gtrsim2.5$ the implied $N_H$ also exceeds $3\times10^{20}$
cm$^{-2}$ even for the two-temperature H atmosphere plus powerlaw
model.

\subsection{Three-Temperature Thermal Spectrum}

The high-energy portion ($\gtrsim$2 keV) of the PSR J0030+0451 X-ray
spectrum could be due to an additional (third) thermal component.
Indeed, for both three-temperature blackbody and H atmosphere models
we obtain statistically acceptable fits.  The best fit parameters for
this model suggest the presence of a very hot and very small ($T_{\rm
  eff}=7$ MK and $R_{\rm eff}=4$ m for a blackbody or $T_{\rm
  eff}=3.4$ MK and $R_{\rm eff}=30$ m for a H atmosphere)
emission region on the neutron star surface. In principle, such a
region could arise due to highly non-uniform magnetospheric heating
across the polar cap surface.

\subsection{Optical-to-$gamma$-ray Spectrum}

%
%  FIGURE 4
%
\begin{figure}[t]
\begin{center}
\includegraphics[width=0.47\textwidth]{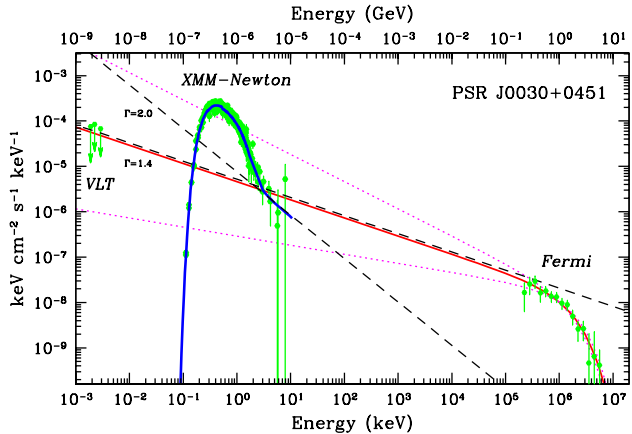}
\caption{Optical-to-$\gamma$-ray spectrum of PSR J0030+0451 showing
  the \textit{XMM-Newton} X-ray data presented in this paper,
  \textit{VLT} optical upper limits \citep{Kop03}, and \textit{Fermi}
  $\gamma$-ray measurements \citep{Abdo09} (\textit{green
    circles}). The solid blue line shows the total absorbed
  two-temperature H atmosphere plus powerlaw model spectrum for
  $\Gamma=2.0$, while the red lines shows the best fit powerlaw model
  (solid) and uncertainties (dotted) for the \textit{Fermi} LAT
  data. The \textit{dashed} lines show the extrapolation of the best
  fit powerlaw spectrum with $\Gamma=2.0$ and $\Gamma=1.4$ of the
  two-temperature H-atmosphere plus powerlaw model to the optical and
  $\gamma$-ray ranges.}
\end{center}
\end{figure}

\citet{Kop03} have conducted deep \textit{VLT} FORS2 BVR imaging of
the field surrounding the radio timing position of PSR J0030+0451. No
plausible optical counterpart was detected down to B $\gtrsim$ 27.3, V
$\gtrsim$ 27.0 and R $\gtrsim$ 27.0 in the immediate vicinity of the
radio pulsar position. In $\gamma$-rays, \citet{Abdo09} have reported
on the \textit{Fermi} LAT detection of pulsations from PSR J0030+0451
above 100 MeV. The $\gamma$-ray spectrum (100 MeV to 10 GeV) is well
described by a powerlaw with a high-energy exponential cutoff with
$\Gamma=1.4\pm0.2$ and a cutoff energy of $E_c=1.7\pm0.4$ GeV.

It is interesting to examine how the X-ray spectrum extrapolates to
optical and $\gamma$-ray energies for the different spectral models
discussed above. Figure 4 shows the optical-to-$\gamma$-ray spectrum
of PSR J0030+0451 based on the available \textit{XMM-Newton},
\textit{VLT}, and \textit{Fermi} observations. The thermal components
seen in the soft X-ray band contribute neglegibly to the optical and
$\gamma$-ray fluxes. On the other hand, for the best fit two-component
H atmosphere thermal plus power-law model, the spectral photon index
of $\Gamma=2$ grossly overestimates the optical flux (by a factor of
$30-40$), violating the VLT upper limits, while it greatly
underestimates the $\gamma$-ray flux (by a factor of $\sim$100).  This
implies that such a powerlaw must break to a flatter powerlaw both in
the UV range and between 10 keV and 100 MeV.

Surprisingly, an extrapolation of the best fit $\gamma$-ray spectrum
with $\Gamma=1.4$ \citep[from][]{Abdo09} is a very close match to the
best fit two-compenent thermal plus non-thermal X-ray spectrum with
$\Gamma=1.4$ (see Table 1). Within the uncertainties of the powerlaw
normalizations as well as the $\sim$20\% effective area uncertainty of
the \textit{Fermi} LAT \citep[see][]{Abdo09}, the two are consistent
with being identical.  This suggests that if the same emission process
is responsible for the hard X-ray tail and the $\gamma$-ray emission
from the pulsar, then the powerlaw index has to be very close to
$\Gamma=1.4$.  However, given the rather large uncertainties in the
two spectra, substantial improvement in the photon statistics in both
energy ranges and detections at intermediate energies (10 keV to 100
MeV) are required to offer more conclusive statements regarding the
relation between the $\gamma$-ray and any non-thermal X-ray emission.

Note that for the Comptonized thermal interpretation of the X-ray
spectrum, one does not expect any contribution of the model components
in the UV/optical nor in the \textit{Fermi} band as the high-energy
cutoff of the Comptonized tail is likely well below the low energy
threshold of \textit{Fermi} LAT ($\sim$100 MeV). Thus, for this model
there is no restriction on the slope of the hard tail.

%
%    FIGURE 5
%
\begin{figure}[t]
\begin{center}
\includegraphics[width=0.47\textwidth]{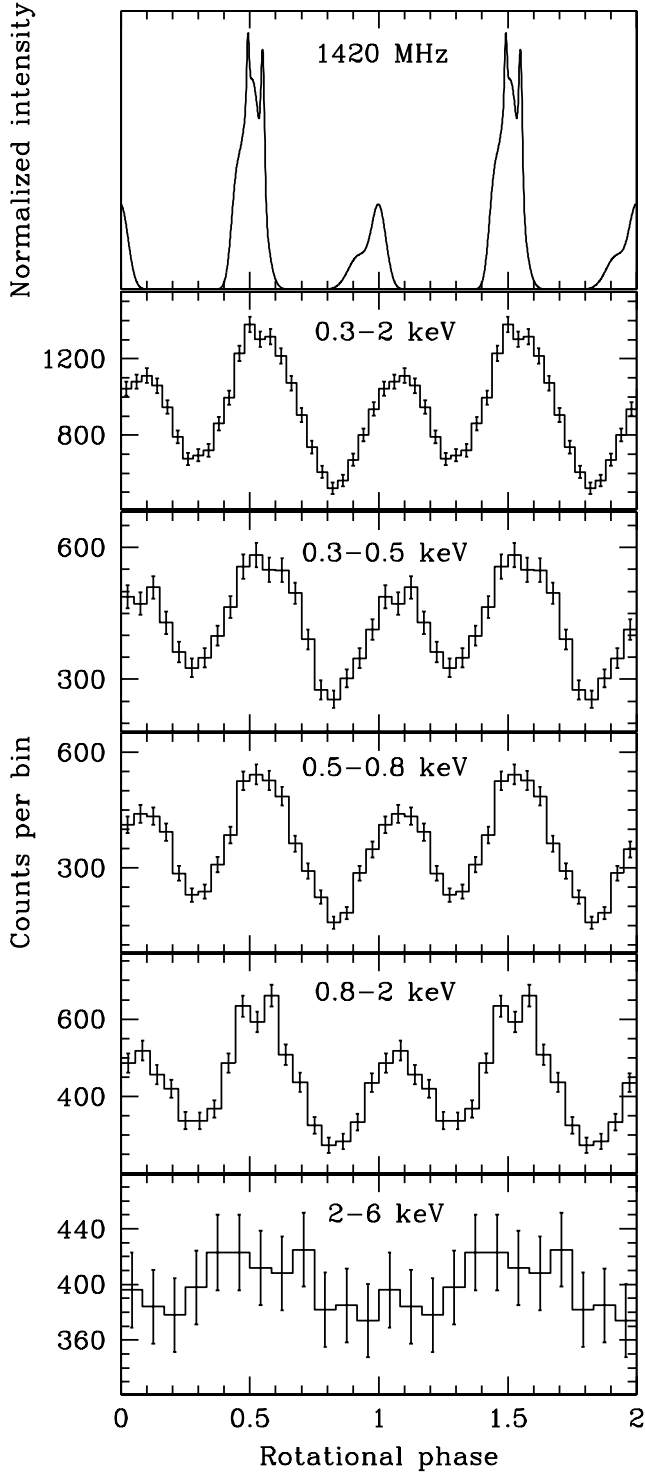}
\end{center}
\caption{\textit{XMM-Newton} EPIC pn X-ray pulse profiles of PSR
  J0030+0451 for different energy bands. The top panel shows the
  template radio pulse profile at 1420 MHz. The choice of phase zero
  and the alignment of the X-ray and radio profiles are arbitrary. Two
  rotational cycles are shown for clarity.}
\end{figure}

\section{TIMING ANALYSIS}

To investigate the rotation-induced modulations of the X-ray flux from
PSR J0300+0451, the photon arrival times extracted from the EPIC pn
source region shown in Figure 2 were first translated to the solar
system barycenter using the SAS {\tt barycen} tool. The arrival times
were subsequently folded at the pulsar spin period using the {\tt
  TEMPO}\footnote{Available at
  http://www.atnf.csiro.au/research/pulsar/tempo/} pulsar timing
package and the radio ephemeris of PSR J0030+0451 from
\citet{Abdo09}. 

As seen from previous X-ray timing studies of this pulsar
\citep{Beck00,Beck02}, the X-ray pulse profile of PSR J0030+0451 is
characterized by two peaks that are significantly broader than
their radio counterparts (Fig.~5).  The substantial improvement
(factor of $\sim$4.5) in photon statistics compared to previous
observations reveals new information regarding the X-ray pulse
morphology. For instance, it is now apparent that the primary pulse is
significantly stronger than the secondary, with a $\sim$13\% larger
peak flux in the 0.3-2 keV range. For thermal polar cap emission this
difference can be naturally produced by a geometric configuration in
which one of the polar caps makes a closer approach to the line of
sight to the observer.  There is marginal evidence for a larger
relative difference between the two pulses for 0.7--2 keV compared to
0.3--0.7 keV, although statistically they are consistent with being
identical.

Based on the lightcurve model discussed in \S6,
the peak-to-peak separation in phase between the two pulses is found
to be $\sim$0.55 as measured from the stronger to the weaker pulse. It
is also evident that the two minima between the pulses are uneven,
with a fractional difference in the minimum flux of $\sim$22\% (for
0.3--2 keV).  For a centered dipole (i.e.~antipodal hot spots),
Doppler boosting and aberration, induced by the rapid stellar
rotation, can be sufficient to account for the apparent assymetry for
larger stellar radii (e.g. $\gtrsim$12 km for $M=1.4$ M$_{\odot}$). On
the other hand, for more compact stars (e.g. $\lesssim$12 km for
$M=1.4$ M$_{\odot}$) it would be necessary to invoke an off-center
dipole in order to reproduce the observed asymmetry in the minima as
well as the phase separation of the two pulses.

To estimate the pulsed fraction of the folded, binned, and
background-subtracted lightcurves we have used two approaches: (i)
from the pulse profile fits discussed in the following section and
(ii) by computing the root-mean-squared pulsed fraction \citep[see,
e.g., Eq.~2 in][for a definition]{Dhillon09}. The former approach
yields 67\%$^{+6\%}_{-6\%}$ (0.3--2 keV), 71\%$^{+9\%}_{-7\%}$
(0.3--0.7 keV), and 63\%$^{+8\%}_{-7\%}$ (0.7--2 keV), whereas using
the latter method we obtain 64\%$^{+4\%}_{-5\%}$,
64\%$^{+6\%}_{-5\%}$, and 63\%$^{+6\%}_{-6\%}$ for the same energy
bands, respectively.  Although these pulsed fractions appear high,
they are fully consistent with a purely thermal origin of the observed
X-rays (as suggested by the spectroscopic analysis in \S4 and as
demonstrated in \S 6) if one considers a light-element atmosphere at
the neutron star surface. Note that the pulse profiles in the 0.3--0.7
and 0.7--2 keV energy bands are consistent with having the same pulsed
fraction. 

Even though the photon statistics are quite limited above 2 keV
and the profile is consistent with a constant flux, the weak
modulations in the 2--6 keV band (bottom panel of Fig.~5) are
suggestive of pulsations similar to those seen at lower energies.
Unfortunately, as a consequence of the relatively high background
level of the EPIC pn data, no useful timing information is present
above $\sim$3 keV, where the hard spectral tail begins to dominate the
emission. Thus, we are unable to place any limits on the pulsed
fraction for 3--10 keV, which could potentially offer clues regarding
the true nature of the hard spectral tail.

%
%    FIGURE 6
%
\begin{figure}[t]
\begin{center}
\includegraphics[width=0.47\textwidth]{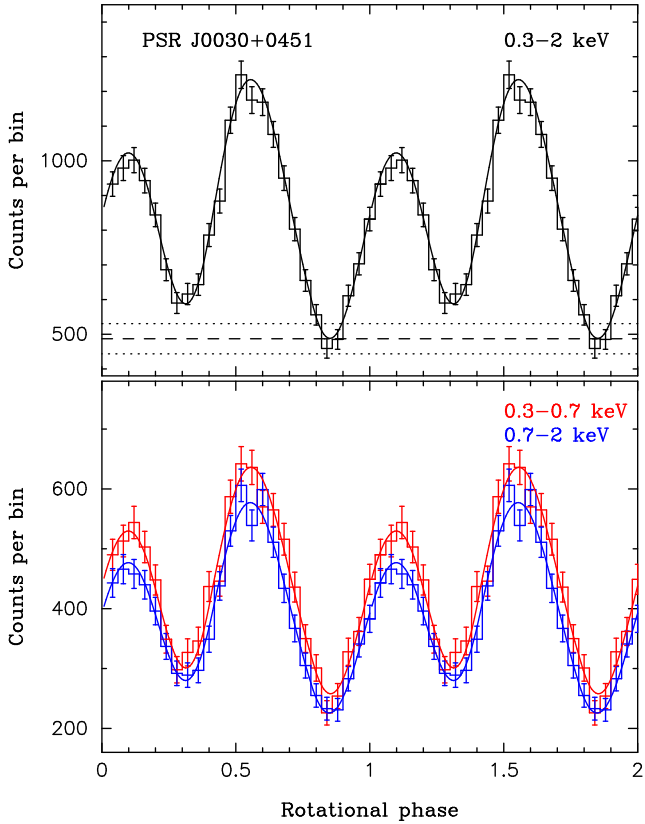}
\caption{\textit{XMM-Newton} EPIC pn pulse profiles of PSR J0030+0451
  in the 0.3--2 keV (\textit{top}) and 0.3--0.7 keV and 0.7--2 keV
  (\textit{bottom}) ranges fitted with a model of a rotating neutron
  star with two-temperature H atmosphere polar caps. The DC level
  (dashed) and uncertainties (dotted) as determined from the fits are
  shown in the top panel. The background level for the 0.3--2 keV
  band is 314 counts per bin for 25 phase bins. See
  text for best fit model parameters.}
\end{center}
\end{figure}

\section{CONSTRAINTS ON THE NEUTRON STAR COMPACTNESS}

The spectrum and pulse profile of PSR J0030+0451 are indicative of
a thermal origin of the observed X-rays. This is of particular
importance since the surface thermal emission can potentially provide
useful information regarding the neutron star compactness through a
measurement of the mass-to-radius ratio ($M/R$).  For this purpose we
employ the model from \citet{Bog07}, which considers a relativistic
rotating compact star with two identical, circular X-ray emitting hot
spots, each (presumably) corresponding to one of the magnetic polar
caps.  The model also incorporates a non-rotating Schwarzschild metric
as a description of the space-time in the vicinity of the star and
includes Doppler boosting and propagation time delays. This relatively
simple formalism is remarkably accurate for spin periods $P\gtrsim3$
ms \citep{Cad07,Mor07}. The NS surface is assumed to be covered by a
non-magnetic, optically-thick H atmosphere.

Before applying the model to the X-ray pulse profiles, it was first
convolved with the appropriate instrument response, while taking into
account the encircled energy fraction and the sky and detector
background.  We fitted the pulse shape by considering the following
parameters: the two temperatures and effective radii of each hot spot
($T_1$, $T_2$, $R_1$, and $R_2$) the two angles $\alpha$ and $\zeta$,
the stellar radius $R$, the offsets of the secondary hot spot from the
antipodal position ($\Delta \alpha$ and $\Delta \phi$), and the phase
$\phi$ of the pulse peak. The hydrogen column density along the line
of sight was fixed at $N_{\rm H}=2 \times 10^{20}$ cm$^{-2}$, while
the distance was set to $D=300$ pc \citep{Lom06}. Varying $N_{\rm H}$
over the range $(1-3)\times10^{20}$ cm$^{-2}$, does not result in
appreciable changes in the most interesting parameters, namely $M/R$,
$\alpha$, and $\zeta$. The same holds true for the parallax distance
uncertainty ($\sim$30\%), which mostly affects the flux normalization,
resulting in a larger uncertainty in the emission area. As $M/R$,
$\alpha$, and $\zeta$ do not determine the flux normalization, the
best fit values of these parameters are virtually insensitive to the
distance uncertainty.  Unless noted otherwise, in our analysis we
assumed a fixed mass of $M=1.4$ M$_{\odot}$ and allow $R$ to vary.
The fit was performed simultaneously in two energy bands (0.3--0.7
keV, with 0.42--0.5 keV excluded as in the spectral fits, and 0.7--2
keV) in order to weaken any degeneracies between spectral parameters
(temperatures and radii) from the geometric and relativistic
parameters, which have no energy dependence. In order to account for
any possible presence of a non-thermal component, we included the
expected number of photons per bin for the best fit powerlaw spectrum
into the uncertainties (assuming an unpulsed or weakly pulsed
contribution). The best fit values for all free parameters were found
by searching the $\chi^2$ hyperspace. To ensure that the obtained
values correspond to the absolute minima of the parameter space, the
fit was repeated for 100 different combinations of initial parameter
values.

The resulting best fit pulse profiles for a H atmosphere model are
presented in Figure 6. Unlike blackbody emission, the atmosphere model
is able to reproduce the pulse shape and large amplitude remarkably
well. The stellar radius is constrained to be $R>10.4$ km or, more
generally, $R/R_S>2.5$ (99.9\% confidence) for all combinations of the
other free parameters. In the case of a blackbody model, good fits to
the X-ray pulse profile require implausably large stellar radii
($\gtrsim$20 km) for a range of assumed massess ($M=1-2$ M$_{\odot}$).
This is in agreement with the results obtained for PSR J0437--4715
\citep{Bog07}, PSR J2124--3358, and using the archival data for PSR
J0030+0451 \citep{Bog08}.

%
%   FIGURE 7
%
\begin{figure}[t]
\begin{center}
\includegraphics[width=0.47\textwidth]{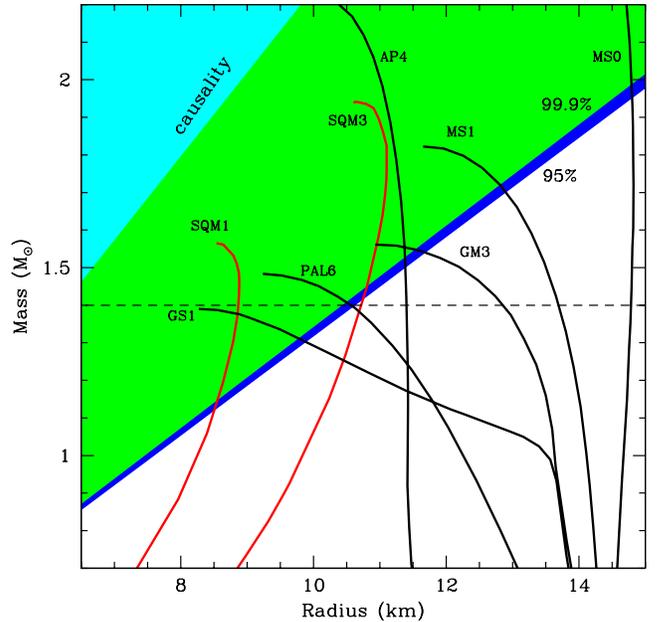}
\caption{The mass-radius plane for neutron stars showing various
  theoretical model tracks. The regions ruled out by the X-ray data
  for PSR J0030+0451 at 95\% and 99.9\% confidence are shown in \textit{blue} and \textit{green}, respectively.}
\end{center}
\end{figure}

Figure 7 shows the $M-R$ plane with a sample of predicted model track
for different NS EOS from \citet[][]{Latt01} and the constraints
obtained for PSR J0030+0451, with the shaded regions corresponding to
regions excluded by the limits on $M/R$. Although the results are
consistent with all EOS if $M$ is unconstrained, it is still possible
to draw some important conclusions. In particular, there is
observational evidence that several recycled pulsars are more massive
than the canonical 1.4 M${\odot}$
\citep{Freire08a,Freire08b,Verb08,Champ08}, as expected from standard
MSP formation theory, which involves accretion of an appreciable
amount of material. If this is true for isolated MSPs such as PSR
J0030+0451 as well, then for $M\gtrsim1.45$ the $M/R$ constraint is
inconsistent with some kaon condensate EOS (e.g.~GS1 for which the
maximum mass consistent with our limits is $\sim$1.3 M$_{\odot}$), and
most quark star EOS (e.g., SQM1--SQM3 in Fig.~7). For $M\gtrsim1.6$
M$_{\odot}$, the limit of $R/R_S>2.5$ rules out all but the stiffest
equations of state (such as MS0--MS2).  Conversely, if this pulsar is
in fact a quark star, then the constraints obtained herein would
require it to be $M\lesssim1.45$ M$_{\odot}$.

The fits to the pulse profile also provide limits on the global
magnetic field and viewing geometry of the pulsar. Specifically, the
pulsar obliquity is constrainted to be in the range
$\alpha=38^{\circ}-90^{\circ}$, implying a substantial misalignment
between the spin and magnetic axes, while the angle between the line
of sight and the spin axis is $\zeta=47^{\circ}-90^{\circ}$. The
relatively broad range of allowed values arises due to the strong
covariance of these angles with the offset in latitude of the
secondary polar cap $\Delta\alpha$, which is weakly constrained to be
in the range $-42^{\circ}-52^{\circ}$. As defined, a positive value of
$\Delta\alpha$ indicates a southward offset from the antipodal
location. The displacement in the longitudinal direction is
constrainted to be $\Delta\phi=-8^{\circ}-46^{\circ}$, consistent with
both a centered and offset dipole. For this parameter, a positive
value corresponds to an offset in the direction of rotation, meaning
that the hot spot leads the antipodal position.

\section{CONCLUSION}

In this paper, we have presented deep \textit{XMM-Newton} observations
of the nearby recycled pulsar PSR J0030+0451. For the first time we
are able to fully characterize the emission spectrum of this isolated
pulsar in the 0.1--10 keV band.  It exhibits striking qualitative
similarities to that of PSR J0437--4715, suggesting that other typical
MSPs such as the nearby J2124--3358 and J1024--0719 \citep{Zavlin06},
as well as the majority of MSPs in 47 Tuc \citep{Bog06a}, whether
binary or isolated, may exhibit the same characteristic spectral
shape over this energy range.

Although the spectral continuum of J0030+0451 can be adequately
described by a variety of plausible models, in all instances the bulk
of detected photons in the soft X-ray range (0.1--2 keV) appear to be
of thermal origin.  Our H atmosphere polar cap model is in fair
agreement with the observed continuum and pulsed emission from PSR
J0030+0451.  As with PSR J0437--4715 \citep{Bog07}, the relatively
large pulsed fractions require the existence of a light-element
atmosphere on the stellar surface and cannot be reproduced by a
blackbody model for realistic NS radii.  By modeling the thermal pulse
shape we are able to place a limit on the allowed stellar radii of
$R>10.4$ km (99.9\% confidence) assuming $M=1.4$ M$_{\odot}$, a
substantial improvement over the constraints obtained from the shorter
archival observations. As demonstrated in \citet{Bog08}, even deeper
X-ray observations of this and other MSPs should lead to stricter
limits on $M/R$, which in turn may firmly rule out entire families of
neutron star EOS.  This is especially true for the growing number of
binary MSPs for which complementary mass measurements from radio pulse
timing are becoming available.

With future observations, it is important to uncover the true nature
of the hard emission ($\gtrsim3$ keV), through a combination of deep
energy-resolved timing and phase-resolved spectroscopy at energies
above 3 keV. Better characterization of the optical and $\gamma$-ray
spectra of PSR J0030+0451 could potentially provide additional information
regarding the origin of this emission.  These endeavors have important
implications for establishing MSP X-ray timing as a vible means for
tight constraints on the neutron star EOS.  The potential utility of
MSP as precision probes of neutron star structure makes them
particularly important targets for upcoming X-ray mission such as the
\textit{International X-ray Observatory}. The great increase in
sensitivity of such a facility would permit detailed observations of a
larger sample of MSPs than currently possible, which could ultimately
lead to definitive insight into the poorly understood properties of
the neutron star interior.

\acknowledgements We would like to thank A.~Lommen for providing the
radio pulse profile of PSR J0030+0451 and the anonymous referee for
offering numerous helpful comments.  The research presented was
funded in part by \textit{XMM-Newton} Guest Observer NASA grant
NNX08AD54G awarded through the Harvard College Observatory. S.~B. is
supported in part by a Lorne Trottier Research Chair Fellowship and a
Canadian Institute for Advanced Research Junior Fellowship.  This work
was based on observations obtained with \textit{XMM-Newton}, an ESA
science mission with instruments and contributions directly funded by
ESA Member States and NASA. The research in this paper has made use of
the NASA Astrophysics Data System (ADS).

%\facilities{XMM-Newton}
Facilities: \textit{XMM-Newton}

\end{document}